\documentclass[aps,prev,twocolumn,preprintnumbers,floatfix,nofootinbib]{revtex4-1}
\pdfoutput=1
\usepackage{graphicx}
\usepackage{epstopdf}
\usepackage{mathrsfs}
\usepackage{amssymb}
\usepackage{verbatim}
\usepackage{color}
\usepackage{multirow}

\newcommand{\beq}{\begin{equation}}
\newcommand{\eeq}{\end{equation}}
\newcommand{\ga}{\lower.7ex\hbox{$\;\stackrel{\textstyle>}{\sim}\;$}}
\newcommand{\la}{\lower.7ex\hbox{$\;\stackrel{\textstyle<}{\sim}\;$}}

\usepackage{hyperref}
\usepackage{amsmath,bm}
\usepackage{physics}

\hypersetup{
    colorlinks = true,
    citecolor = {blue},
    linkcolor = {blue},
    urlcolor = {blue},
}

\begin{document}

\def\jcap{\ref@jnl{J. Cosmology Astropart. Phys.}}

\vspace{0.5cm}
\title{A Unified No-Scale Model of Modulus Fixing, Inflation, Supersymmetry Breaking \\ and Dark Energy}

\author{John~Ellis$^{a}$}
\author{Dimitri~V.~Nanopoulos$^{b}$}
\author{Keith~A.~Olive$^{c}$}
\author{Sarunas~Verner$^{c}$}

\affiliation{
$^a$Theoretical Particle Physics and Cosmology Group, Department of
  Physics, King's~College~London, London WC2R 2LS, United Kingdom;\\
Theoretical Physics Department, CERN, CH-1211 Geneva 23, Switzerland;\\
NICPB, R\"avala pst. 10, 10143 Tallinn, Estonia}
\affiliation{$^b$George P. and Cynthia W. Mitchell Institute for Fundamental
 Physics and Astronomy, Texas A\&M University, College Station, TX
 77843, USA; \\
 Astroparticle Physics Group, Houston Advanced Research Center (HARC),
 Mitchell Campus, Woodlands, TX 77381, USA; \\
Academy of Athens, Division of Natural Sciences,
Athens 10679, Greece,}
\affiliation{$^c$William I. Fine Theoretical Physics Institute, School of
 Physics and Astronomy, University of Minnesota, Minneapolis, MN 55455,
 USA}

\begin{abstract}

We present a minimal SU(2,1)/SU(2)$\times$U(1) no-scale model that unifies modulus fixing, Starobinsky-like inflation, an adjustable scale for supersymmetry
breaking and the possibility of a small cosmological constant, a.k.a. dark energy.\\
\begin{center}
{\tt KCL-PH-TH/2019-24, CERN-TH-2019-024, ACT-01-19, MI-TH-1917, \tt UMN-TH-3820/19, FTPI-MINN-19/11} 
\end{center}

\end{abstract}

\maketitle


Physics contains many hierarchies of mass scales, starting from the Planck scale $M_P \sim 10^{19}$~GeV at which the effects of
quantum gravity must become important, through the energy scale of cosmological inflation, which is $\lesssim 10^{13}$~GeV,
through the electroweak scale $\sim 100$~GeV down to the energy scale
of dark energy, a.k.a. the cosmological constant, which is $\sim 10^{-3}$~eV. What are the origins of these hierarchies, 
and how can they be stabilized in a natural way despite the depredations of quantum corrections?
Diverse origins have been proposed, and this paper does not claim any progress in elucidating this aspect of the hierarchies.
Instead, we focus on the question of how they can be accommodated within a simple framework that incorporates a
mechanism for stabilizing hierarchies of mass scales.

That framework is provided by supersymmetry, which could stabilize the electroweak hierarchy if the supersymmetry-breaking
scale is $\lesssim 1$~TeV  \cite{Maiani:1979cx}, and could also stabilize the parameters of an inflationary scalar potential at some scale $\ll M_P$ \cite{cries}.
On the other hand, simple supersymmetry is insufficient by itself to render natural the small magnitude of the cosmological
constant, and it would need to be supplemented by dynamical mechanisms to generate the hierarchies of mass scales.
In the context of cosmology, supersymmetry must be combined with general relativity within some form of supergravity
theory \cite{Nilles:1983ge}. For this we advocate no-scale supergravity \cite{no-scale,LN}, since it does not have any of the deep ${\cal O}(M_P^4)$ anti-de Sitter
(AdS) `holes' in the effective potential that are endemic in other supergravity theories with matter fields.
Moreover, no-scale supergravity appears generically in compactifications of string theory~\cite{Witten}, which we regard as the most promising candidate for a quantum theory of gravity. One may anticipate that string theory is 
the UV completion of the model presented below, though our model does require specific knowledge of string theory
other than this recognition that no-scale supergravitiy is a generic consequence of string models~\cite{Witten}. 

No-scale supergravity has been shown to yield Starobinsky-like models of inflation~\cite{Staro}, under suitable conditions on the
theoretical parameters~\cite{ENO6}, and we have recently characterized general conditions under which de Sitter (dS) vacua can be
accommodated within no-scale supergravity~\cite{enno}. Upgrading such models to something resembling the Standard Model (SM)
in a more realistic way requires a deeper discussion on how matter fields should be incorporated, that should also include
a mechanism for supersymmetry breaking. Previously, this has often been done by invoking some variant of the Polonyi
model in which supersymmetry breaking originates dynamically within a hidden sector~\cite{pol,bfs,ego,egno4,rking}.

In this paper we build upon~\cite{enno,ENOV}, generalizing the characterization of de Sitter (dS) 
no-scale supergravity models with SU(2,1)/SU(2)$\times$U(1) K\"ahler manifolds. Without extending the field content,
we introduce a mechanism that allows for 
a transition between two dS vacua, one that can accommodate Starobinsky-like inflation and one with an amount of
vacuum energy that could be very small, like the present cosmological constant (dark energy), 
without invoking any external
mechanism such as uplifting by fibres~\cite{KKLT}. 
As we show, this class of models also allows for supersymmetry breaking with a
magnitude suitable for stabilizing the electroweak hierarchy, without invoking any hidden sector {\it \`a la} Polonyi. Additionally, a mechanism proposed previously~\cite{EKN3,ENO7} can be used to fix the compactification modulus of the SU(2,1)/SU(2)$\times$U(1) model.

We recall that a supergravity theory is described by a K\"ahler function $G \equiv K + \ln W + \ln W^{\dagger}$, 
where $K$ is a Hermitian K\"ahler potential and $W$ is a holomorphic superpotential. The scalar kinetic terms of the Lagrangian are given by
$\mathcal{L}_{kin} = K^{i j^{*}} \partial_{\mu} \Phi_i \partial^{\mu} \Phi_{j}^*$,
where $K^{i j^{*}} \equiv \partial^2{K}/\partial_{\mu} \Phi_i \partial^{\mu}\Phi_{j}^*$ is the K\"ahler metric,
and the effective scalar potential is given by
\begin{equation}
V = e^{G} \left[ \pdv{G}{\Phi_i} K_{i j^*} \pdv{G}{\Phi_{j}^{*}} - 3 \right] \, ,
\label{effpot}
\end{equation}
where $K_{ij^*}$ is the inverse of the K\"ahler metric.

Our model is characterized by the following K\"ahler potential:
\begin{eqnarray}
K  & = & - 3 \, \alpha \, \ln \left[ T + T^\dagger - \frac{\phi \phi^\dagger}{3} -  \frac{X^i X_i^\dagger}{3}  \right. \nonumber \\
& + &  \left. b (T+T^\dagger-2d)^4 + c (T-T^\dagger)^4  \right] \, ,
\label{K}
\end{eqnarray}
where $\alpha$ is a curvature parameter
(hereafter we set $d = 1/2$ for definiteness) and the superpotential $W$ can be written as follows:
\begin{equation}
W \; = \;  W_I (T, \phi) + W_{SM} (X,\phi) + W_{dS} (T, \phi)
\label{W}
\end{equation}
where $W_I$ is responsible for inflation, $W_{SM}$ contains Standard Model interactions (possibly with the inflaton for reheating), and $W_{ds}$ will provide both supersymmetry breaking and dark energy.
More specifically, these are
\begin{equation}
W_I \; = \;  M  \left( \frac{\phi^2}{2} - \frac{\phi^3 }{3\sqrt{3}}\right)\,  ,
\label{WIWSM1}
\end{equation}
\begin{equation}
W_{SM} \; = \; \mu X^i X^j + \lambda X^i X^j X^k  + y \phi X^i X^j  \, ,
\label{WIWSM2}
\end{equation}
and
\begin{equation}
W_{dS} \; = \; a_1 \left(2T - \frac{\phi^2}{3}  \right)^{n_{-}}  - a_2 \left(2T - \frac{\phi^2}{3}  \right)^{n_{+}} \, .
\label{WDS}
\end{equation}
Eqs. (\ref{K}) - (\ref{WDS}) fully define our model. While inflation based on Eq. (\ref{WIWSM1}) was first introduced in 
\cite{ENO6}, combining this with $W_{SM}$ and $W_{dS}$ in Eq. (\ref{W}) is the key novel feature in the model
considered below. In particular, adding $W_{dS}$ to $W_I$ preserves Starobinky-like inflationary evolution
while breaking supersymmetry and leaving residual vacuum energy suitable for dark energy today.
This is accomplished in a remarkably simple form without the necessity of addition field content.
We now preview the interpretations of these expressions, before discussing them in more detail in the bulk of the paper.

The complex field $T$ can be interpreted as a volume modulus of compactification, and $\phi$ as another gauge-singlet modulus acting as the inflaton.
Together they parameterize the no-scale SU(2,1)/SU(2)$\times$U(1) coset manifold \cite{ENOV,clear}, while $X^i$ represent gauge-non-singlet matter fields such as those
appear in the SM. The quartic terms in (\ref{K}) fix $T$~\cite{EKN3,ENO7}, $W_I$ in (\ref{WIWSM1}) fixes the inflaton $\phi$
and enables Starobinsky-like inflation with an energy scale ${\cal O}(M)$ (other forms for $W_I$ are also possible: see~\cite{ENO7,ENOV}), and
$W_{SM}$ in (\ref{WIWSM2}) contains bilinear and trilinear terms of the generic forms appearing in SM-like superpotentials
as well as a coupling of the inflaton to matter to allow for reheating.
The novel terms in (\ref{W}) are those in (\ref{WDS}) with coefficients $a_{1,2}$, which have functional forms that are holomorphic
versions of the corresponding Hermitian terms of the gauge singlets in the K\"ahler potential (\ref{K}). 
Taken alone, $W_{dS}$ leads to a de Sitter solution for all real values of $\phi$ and $T$.
The couplings $a_1$ and $a_2$ determine the magnitudes of
soft supersymmetry breaking and the cosmological constant, which are ${\cal O}(a_1 - a_2)$ and ${\cal O}(a_1 a_2)$, respectively.	
Choosing $a_1 = {\cal O}(10^{-16})$ and $a_2 \ll a_1$ (or {\it vice versa}) would yield a cosmological constant and soft supersymmetry
breaking of the desired magnitudes. 
No-scale supergravity theories derived from string models have, in general, additional 
moduli such as the dilaton and complex structure moduli. For simplicity, we
assume that 
these are already fixed at scales above the inflationary scale.
 Throughout, we work in units of the reduced Planck mass, $M_P = 1/(8\pi G_N) \approx 2.4 \times 10^{18}$ GeV.

We consider now in more detail the dS/dark energy sector $W_{dS}$~(\ref{WDS}).
Constructions of dS vacua  with multiple moduli in SU(1, 1)/U(1) x U(1) no-scale supergravity were discussed in~\cite{enno},
and can be extended to general SU(N, 1)/SU(N) x U(1) K\"ahler coset manifolds via the K\"ahler potential
$K = - 3 \, \alpha \ln \left[ T + T^{\dagger} - \frac{\phi_i \phi_{i}^\dagger}{3} \right]$.
We find that dS vacua solutions can be obtained from a superpotential $W_{dS}$ of the form (\ref{WDS})
with $\phi \to \phi_i$ and exponents given by $n_{\pm} = \frac{3}{2} \left( \alpha \pm \sqrt{\alpha} \right)$~\cite{EKN1,rs,enno}. 
Holomorphy of the superpotential requires $\alpha \geq 1$, and
the no-scale case $K$ corresponds to $\alpha = 1$~\cite{ENO9}.
We assume that the imaginary parts of the moduli fields are fixed to 
$\langle {\rm Im}~T \rangle = 0$ and $\langle {\rm Im}~\phi_i \rangle = 0$, which can be realized by introducing higher-order terms in the K\"ahler potential 
such as those in the second line of (\ref{K}),
as was shown in~\cite{EKN3,ENO7}, or (in some cases) by the dynamics of the potential. Specializing to the SU(2, 1)/SU(2) x U(1) no-scale
case and inserting the expressions~(\ref{K}) and~(\ref{WDS}) into~(\ref{effpot}), 
we find the following effective scalar potential at the minimum:
\begin{equation}
V = 12 \, a_1 \, a_2 \, ,
\label{pot1}
\end{equation}
for all values of ${\rm Re} \,  T$ and ${\rm Re} \,  \phi$, which 
corresponds to a de Sitter vacuum when $a_1$ and $a_2$ are of the same sign. Thus the dS/dark energy superpotential $W_{dS}$ yields a cosmological constant~(\ref{pot1})
following the end of inflation.\\

Combining $W_{dS}$ with a suitable inflationary superpotential $W_{I}$, we can incorporate soft supersymmetry breaking without adding an
additional Polonyi sector~\cite{pol,bfs} or introducing other possible dynamics~\cite{KKLT}. To this end,
we consider an inflationary superpotential $W_I$  that vanishes when the complex scalar fields $T$ and $\phi_i$ are fixed at the potential minimum, i.e.,
we do not induce supersymmetry breaking through $W_I$, which typically has a mass scale of the order of the inflationary scale $\sim 10^{13}$ GeV.
When the volume modulus $T$ is stabilized through the quartic terms in Eq. (\ref{K}) so that ${\rm Re} \,  T = 1/2$ and ${\rm Im} \,  T = 0$,  the inflaton is 
stabilized so that ${\rm Im} \,  \phi = 0$ throughout inflation and ${\rm Re} \,  \phi = 0$ at the end of inflation.
Supersymmetry breaking is generated through an $F$-term for $T$, which is given (for $\alpha = 1$) by
\begin{eqnarray}
F_T & = &  - e^{G/2} K_{ij^*} G^j = - m_{3/2} (K_T + W_T/W)/3  \nonumber \\
& = & (a_1 + a_2) \ne 0
\end{eqnarray}
at the minimum, and is independent of the inflationary scale $M$. The gravitino mass is simply given by $m_{3/2} = a_1 - a_2$~\cite{f2}.

Supersymmetry breaking with a Minkowski vacuum would be obtained if either $a_1$ or $a_2$ vanishes, but we are interested here in models with $a_{1, 2} \ne 0$.
Specifically, we choose $a_1 - a_2 = \mathcal{O} (10^{-16})$ in order that the gravitino mass be 
$ \mathcal{O} (1)$~TeV~\cite{f3}. If we also choose
$a_2 = \mathcal{O}(10^{-104})$, we would obtain a value of the dark energy density (cosmological constant) comparable to the present value, $\mathcal{O} (10^{-120})$.
However, this is not the appropriate choice, since we expect other contributions
to the present vacuum energy density, specifically from electroweak gauge symmetry breaking and confinement in QCD, which are estimated to be
$\mathcal{O} (m_W^4) \sim \mathcal{O} (10^{-68})$ and $\mathcal{O} (\Lambda_{QCD}^4) \sim \mathcal{O} (10^{-80})$, respectively.
These can be accommodated together with the present value of the dark energy density by choosing $a_1 = \mathcal{O} (10^{-16})$
and $a_2 = \mathcal{O} (10^{-52})$ with the values finely-tuned to cancel the electroweak and QCD contributions
so that the net value of the dark energy is $\mathcal{O} (10^{-120})$.
It is also possible that  $a_2 = \mathcal{O} (10^{-20})$ if the residual vacuum density is canceled by
a grand unified theory (GUT) phase transition of order $m_{3/2}^2 M_{GUT}^2$ \cite{EGNNO}.
The couplings $a_1$ and $a_2$ are radiatively stable, but the required
fine-tuning is inelegant. We have no suggestion on how it may be achieved dynamically in a natural way, but it does provide an effective framework for the different relevant mass scales
without additional fields or resorting to uplifting with additional string dynamics.

One expects that the mass of the inflaton may be of order $M \sim \mathcal{O} (10^{-5}) $, in which case
we can safely ignore the mixing terms between $a_{1,2}$ and $M$ during inflation because $a_2 \ll a_1 \ll M$,
so the constants $a_{1,2}$ do not affect the slow-roll inflationary dynamics. As was shown in~\cite{ENO6,ENO7}, the Starobinsky inflationary potential can be recovered in no-scale  SU(2,1)/SU(2) x U(1) supergravity
from the Wess-Zumino-like superpotential $W_I$ in (\ref{WIWSM1}), and we recall briefly some of the results. 
For simplicity, we consider the K\"ahler potential~(\ref{K}) with two moduli fields $\phi$ and $T$, and set the curvature parameter to $\alpha = 1$.
We assume that the scalar field $T$ is fixed by the quartic terms in Eq. (\ref{K}), acquiring a vev $\langle T \rangle = 1/2$.
The couplings $b$ and $c$ are expected to both be $\gg 1$, corresponding to inverse
mass scales $b, c \propto 1/\Lambda_T^2$, with $\Lambda_T \ll M_P$, similar to strong stabilization in some Polonyi models
\cite{dine,Dudas:2006gr,klor,dlmmo,eioy,nataya,ADinf,ego,egno4}. 
In the limit $a_2 \to 0$, if $b = c$, the two real fields associated with $T$ acquire the same mass. 
However, in the absence of supersymmetry breaking both remain massless \cite{ENO7}, 
and 
only acquire a mass when supersymmetry is broken, ${m_{Re \, T}} = 4 \sqrt{3b} a_1 \simeq 4 \sqrt{3b} m_{3/2}$, 
${m_{Im \, T}} = 4 \sqrt{3c} a_1 \simeq 4 \sqrt{3c} m_{3/2}$.

Although the mass of $T$ is significantly below the inflaton mass,
there is no Polonyi-like problem \cite{polprob} associated with $T$. 
As in the strongly-stabilized Polonyi system, the dominant decay mode 
for $T$ is into a gravitino pair \cite{ego}, with a decay rate proportional to
$m_{3/2}^3 M_P^3/\Lambda_T^5$~\cite{f4}.
The problem of entropy production is easily evaded here. 
For $\Lambda_T \lesssim 10^{-4}$, the modulus decays before the inflaton
and plays little role in subsequent reheating processes. 
As a result, there is no additional constraint from
the overproduction of gravitinos (and ultimately the lightest supersymmetric particle). 
Since our stabilization
term in the K\"ahler potential should be thought of as an effective interaction,
consistency would require $\Lambda_T > {m_T}$ or $\Lambda_T > (48)^{1/4} (m_{3/2} M_P)^{1/2}$.
Thus there is a substantial range of values for $\Lambda_T$ for which there is no moduli problem:
see \cite{ego} for further details. It is interesting that $\Lambda_T = M$ lies within the allowed range.

With the modulus fixed at $\langle T \rangle = 1/2$ and ${\rm Im}\, \phi = 0$ 
(which minimizes the potential), so that Re~$\phi = \phi$, we can write the scalar potential as
 \begin{eqnarray}
V &  = & 12 a_1 a_2 + 12 a_2 M   \left(\frac{\phi^2}{2} - \frac{\phi^3}{3 \sqrt{3}} \right)  \nonumber \\
& + & 3 M^2 \left( \frac{\phi}{\sqrt{3} + \phi } \right)^2  \, .
\label{phipot}
\end{eqnarray}
The first term is the cosmological constant, and the second term is a perturbation of the inflaton potential that
 has a negligible effect on the inflationary dynamics, because $a_2 \ll M$. 
 We then make
 the following field redefinition to obtain a canonical kinetic term for ${\rm Re} \,  \phi$:
\begin{equation}
{\rm Re} \, \phi =   \sqrt{3} \tanh (x/\sqrt{6} ) \, ,
\label{arctanh}
\end{equation}
With this field redefinition and dropping terms proportional to $a_1$ and $a_2$,
we obtain the Starobinsky potential along the ${\rm Re} \,  \phi$ direction:
\beq 
V =  \frac{3}{4} M^2 \left(1- e^{-\sqrt{2/3}x} \right)^2 \, .
\label{r2pot}
\eeq
The first two terms in Eq. (\ref{phipot})
can be written as
\beq
\Delta V = \Lambda + 6 a_2 M \tanh^2 (x/\sqrt{6}) \left( 3 - 2 \tanh (x/\sqrt{6}) \right),
\label{deltapot}
\eeq
where we have defined the cosmological constant $\Lambda = 12 a_1 a_2$. 
We note that, at large $x$, $\Delta V$ adds a relatively small amount $6 a_2 M$ to the Starobinsky
plateau value of $(3/4) M^2$.

\begin{figure}[ht!]
\centering
\includegraphics[scale=0.55]{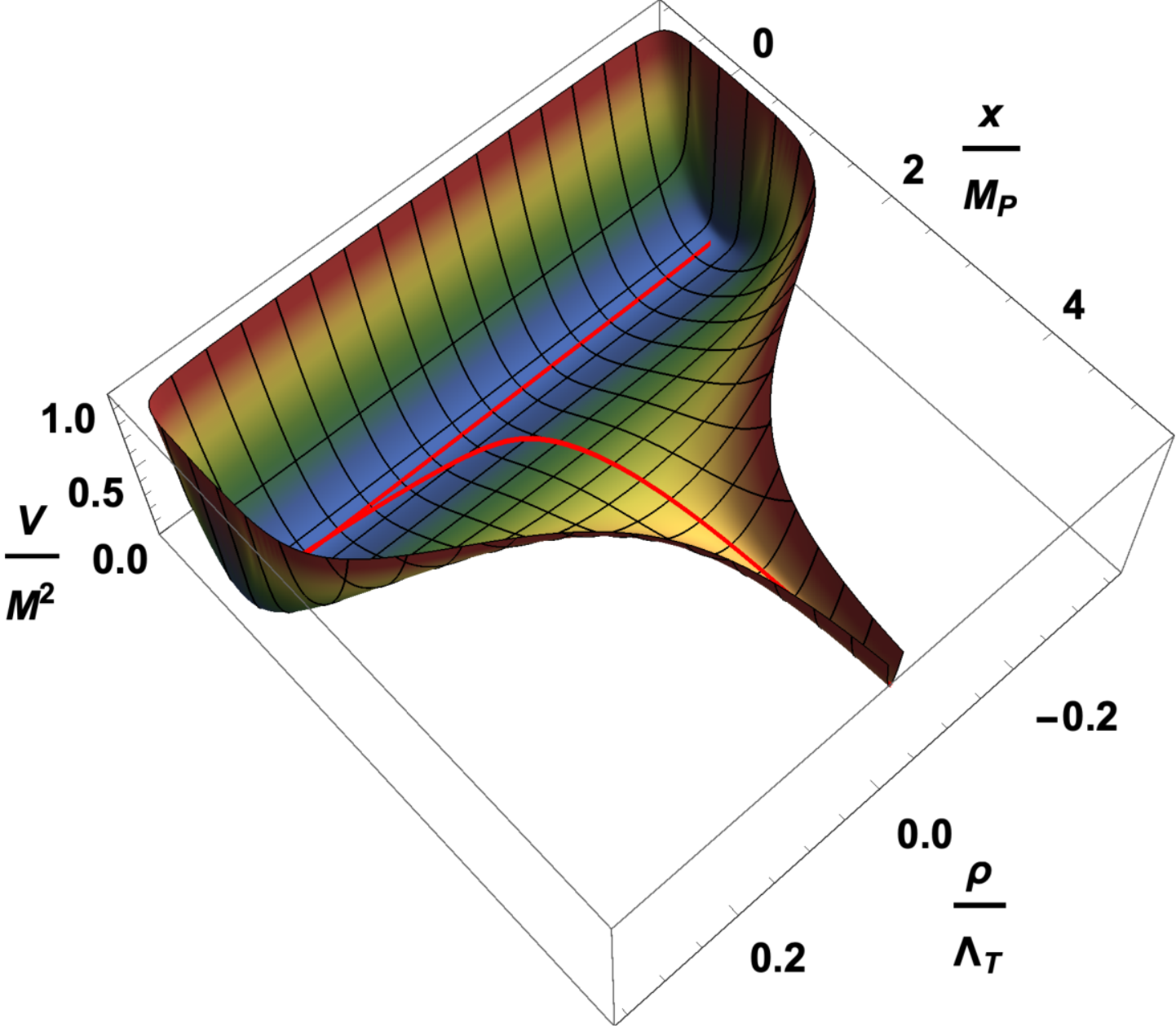}
\caption{\it Realization of the Starobinsky inflationary potential with the initial values of $x(0) = 5.35$, $\rho(0) = 0$. 
The amplitude of the oscillation and the location of the barrier is given by $\langle \rho/\Lambda_T \rangle_{Min} \sim 0.34$.  }
\label{wzplot}
\end{figure}

To visualize slow-roll inflation in the Re~$\phi$ - Re~$T$ plane,
we use the following field redefinition:
\begin{equation}
{\rm Re} \, T = \frac{1}{2} e^{- \sqrt{\frac{2}{3}} \rho},
\label{realT}
\end{equation}
together with the field redefinition ${\rm Re} \, \phi \to x$~(\ref{arctanh}). The scalar potential acquires a complicated form in terms of $(x, \rho)$
that we do not display here, which reduces to the form $V + \Delta V$ given by~(\ref{r2pot}) and~(\ref{deltapot}) when $\rho = 0$. We assume
that the number of e-folds until inflation ends is $N = 55$, which is realized with the nominal choice of $x = 5.35$ and $\rho = 0$, 
yielding the tensor-to-scalar ratio $r = 0.0035$ and the spectral tilt $n_s = 0.965$~\cite{planck18,egno3}. 
The potential in the $x$ - $\rho$ plane is shown in Fig. \ref{wzplot}. The field $x$ exits the dS plateau and rolls down towards a potential barrier on the left, 
located at $x = 0, \rho \approx 0.34 \, \Lambda_T$. Then the field $\rho$ evolves slowly towards the global minimum, located at $x = 0, \rho = 0$, 
and starts spiraling about the minimum with initial amplitude $\langle \rho \rangle_{Min} \approx 0.34 \, \Lambda_T $ and $\langle x \rangle \approx M_P$
until $\rho$ (or $T$) decay which occurs well before reheating when the inflaton decays.

Finally, we consider the full model with SM-like fields characterized by the K\"ahler function~(\ref{K}), 
and the full superpotential $W = W_{dS} + W_{I} + W_{SM}$~(\ref{W}). After the modulus acquires its vacuum expectation value, 
$\langle T \rangle = \frac{1}{2}$, and as the inflaton settles to its minimum, the reheating process begins.
The coupling $y \phi X^i X^j$ provides a decay channel for the inflaton and, assuming
instantaneous reheating, the reheat temperature is $T_R \sim y (M M_P)^{1/2}$.

At the minimum, $\langle T \rangle = 1/2$ and 
$\langle \phi \rangle = 0$, and the potential for the SM-like fields $X^i$ can be written as follows in the limit $M_P \to \infty$:
\begin{equation}
V_{SM} = \sum_i |W_{X^i} |^2 + 2 a_2
\mu \left(X^i X^j + h.c. \right)  + 12 a_1 a_2 \, ,
\label{VSM}
\end{equation}
where $W_{X^i} \equiv \partial W / \partial X^i$. The first term in (\ref{VSM}) is simply a (global) supersymmetric potential term for the 
Standard Model fields $X^i$ in no-scale supergravity.
The second term is generated from supersymmetry breaking
and appears as an effective supersymmetry-breaking bilinear term, $B_0 =  2 a_2$.
The third term is, again, our cosmological constant. 
Gaugino masses can be generated if the gauge kinetic function, $f_{\alpha \beta}$
is a function of $T$, so that $m_{1/2} \simeq  F_T |f_T^\dagger/f|/2 \propto m_{3/2}$.

As an alternative to (\ref{K}), one could also consider the case of twisted matter fields,
where the kinetic term $X X^\dagger$ is taken out of the logarithm in 
Eq, (\ref{K}). In such a model, additional soft mass terms are generated as the scalar potential becomes
\begin{eqnarray}
V_{SM} &  = &  \sum_i |W_{X^i} |^2 + (a_1 - a_2)^2 \sum_i  |X^i|^2 \nonumber \\
&+& ( 2 a_1 + 4 a_2)\mu (X^i X^j + h.c.)  \nonumber \\
& + & 3 (a_1 + a_2)\left[  \lambda X^i X^j X^k +  y \phi X^i X^j + h.c. \right]  \nonumber \\
& + & 12 a_1 a_2 \, ,
\end{eqnarray}
where we see soft scalar masses with $m_0 = (a_1 - a_2) = m_{3/2}$,
a bilinear term, $B_0 = (2 a_1 + 4 a_2)$ and trilinear
$A$-terms, $A_0 = 3 (a_1 + a_2)$. 

We have outlined in this paper a simple no-scale supergravity framework for sub-Planckian physics capable of including
modulus fixing, Starobinsky-like inflation at a scale $\mathcal{O}(10^{13})$~GeV, supersymmetry breaking 
at a scale $\mathcal{O}(10^{3})$~GeV, and a small positive cosmological constant (dark energy density).
This model should not be regarded as fundamental, but rather as an effective field theory that should, we believe,
ultimately be derived from a suitable variant of string theory.
In a more complete study of the dynamics of this model, the renormalization-group
evolution of the supersymmetry-breaking terms would be able to drive the effective Higgs mass-squared negative,
triggering electroweak symmetry breaking~\cite{Ellis:1983sf} and the corresponding change in the vacuum energy density. As mentioned earlier,
the parameters $a_{1,2}$ should be chosen such that the dark energy density takes its physical value $\mathcal{O}(10^{-120})$
after this contribution is included. Thus the electroweak scale could be generated dynamically in this framework,
though we have no new suggestions to offer concerning the origins of the inflationary, supersymmetry-breaking or
dark energy scales. Finally, we note that it would be interesting to explore the extension of this scenario to
include grand unification~\cite{EKN2} and related issues such as neutrino masses and mixing~\cite{EGNNO}. However, such topics lie
beyond the scope of this paper.

\subsection*{Acknowledgements}
We would like to thank K. Kaneta for useful discussions.
The work of JE was supported in part by the United Kingdom STFC Grant
ST/P000258/1, and in part by the Estonian Research Council via a
Mobilitas Pluss grant. The work of DVN was supported in part by the DOE
grant DE-FG02-13ER42020 and in part by the Alexander~S.~Onassis Public
Benefit Foundation. The work of KAO was
supported in part by DOE grant DE-SC0011842 at the University of
Minnesota.


\begin{thebibliography}{9}
\bibitem{clear}
The part of the model that has SU(2,1)/SU(2) $\times$U(1) symmetry  
is found in Eq. (\ref{K}) when $b = c =0$ and the SM fields are ignored.
If the SM fields are included the symmetry is upgraded to an SU(N,1)/SU(N) $\times$ U(1) symmetry.
\bibitem{f2}
Note that the common relation $m_{3/2} = F/\sqrt{3}$ is not satisfied here as do not have a Minkowski vacuum.
Recalling that we can write $V = G^{ij} F_i F_j - 3 m_{3/2}^2$, we easily recover Eq. (\ref{pot1}) at the minimum.
\bibitem{f3}
We note that the roles and values of $a_1$ and $a_2$ are largely interchangeable.
\bibitem{f4}
Up to a factor of 216, the decay rate for 
$T$ is the same as the decay rate for a strongly-stabilized Polonyi field.
The limit on $\Lambda_T$ will in general depend on $m_\phi, m_{3/2}$ and $y$.
\end{thebibliography}
\end{document}